\documentclass[aps,pra,twocolumn,10pt]{revtex4-2}

\usepackage{xcolor}

\usepackage{epsfig,amsmath}
\usepackage{subfigure}
\usepackage{graphicx}
\usepackage{dcolumn}
\usepackage{stmaryrd}
\usepackage{mathrsfs}
\usepackage{pifont}
\usepackage{amsthm}
\usepackage{amssymb}
\usepackage{bm}
\usepackage{braket}
\usepackage{adjustbox}
\usepackage{latexsym}
\usepackage{color}
\usepackage{multirow}
\usepackage{xcolor}
\usepackage{babel}
\usepackage[ruled]{algorithm2e}
\usepackage{verbatim}
\usepackage{enumerate}
\usepackage{placeins} 
\usepackage{booktabs}

\usepackage{dsfont}
\usepackage{float} 
\usepackage{amsmath,amssymb}
\usepackage[utf8]{inputenc}
\usepackage{tabularx} 
\usepackage{url}           
\usepackage{hyperref}
\usepackage{caption}
\begin{document}

\title{Higher-Order Magnus Corrections to Leakage Elimination Operator-Based Quantum Control Pulse Design in the Low-Frequency Regime}
\author{Hang Yu\textsuperscript{1}, Kai-Yu Yuan\textsuperscript{2}, Feng-Hua Ren\textsuperscript{3}, Zhao-Ming Wang\textsuperscript{1,4,5}}
\email{Contact author:wangzhaoming@ouc.edu.cn}

\address{$^{1}$ College of Physics and Optoelectronic Engineering, Ocean University of China, Qingdao 266100, China}
\address{$^{2}$ China Mobile (Suzhou) Software Technology Co., Ltd, Suzhou 215163, China}
\address{$^{3}$ School of Information Management and School of Artificial Intelligence, Qingdao University of Technology, Qingdao 266520, China}
\address{$^{4}$	Engineering Research Center of Advanced Marine Physical Instruments and Equipment of Ministry of Education, Qingdao 266100, China}
\address{$^{5}$ Qingdao Key Laboratory of Optoelectronics, 	Ocean University of China, Qingdao 266100, China}

\begin{abstract}
Effective quantum control is essential for quantum computing and information processing, yet level leakage remains a bottleneck limiting control fidelity. The leakage elimination operator (LEO) approach is highly effective at suppressing leakage from a target subspace to other subspaces. Analytical control conditions under the high-frequency driving limit have been derived via the Feshbach PQ partitioning technique. However, the driving intensity is subject to fundamental physical bounds, and low-frequency driving is experimentally more feasible. In this work, we reformulate LEO within the Magnus framework and extend it beyond the high-frequency limit. We show that the analytical control conditions obtained from the Feshbach PQ partitioning technique are equivalent to the zeroth-order Magnus expansion. We then analyze the higher-order Magnus corrections. By minimizing the leakage, we obtain a modified pulse condition suitable for the low-frequency regime. As an illustrative example, we study near-perfect state transfer through a one-dimensional spin chain under LEO control. We find  that the second-order Magnus correction improves the transmission fidelity at low-frequency driving, while the improvement becomes negligible at high frequencies. Our results provide an effective route for designing control pulses in the low-frequency regime, with potential applications in superconducting qubits, trapped ions, and solid-state spin systems.
\end{abstract}
\maketitle

\section{Introduction}
Quantum control requires the high precision manipulation of the system dynamics~\cite{NP}, with broad applications in atom~\cite{DE}, optics~\cite{TS}, and solid state physics~\cite{TO}. In quantum information processing, pulses are often used to control the qubits, such as radio-frequency pulses in superconducting quantum processors~\cite{Zeissler2023,ZhangZewen2025} and shaped laser pulses in Rydberg atoms~\cite{CaiP2019,ZhangBin2014}. To achieve high-precision quantum control, various numerical~\cite{11201021,Bhole2018,Kro} and analytical~\cite{VanDamme2017,WAEST} methods have been developed for pulse design. Dynamical decoupling~\cite{VD,VU} is one of the effective control strategies and has been widely applied in quantum control field, where the leakage elimination operator (LEO) approach~\cite{LEO,WE} is particularly effective in suppressing leakage from a subspace to others~\cite{BU,AF,DQ}. 

In the LEO scheme, the control function can be chosen arbitrarily~\cite{Wunotes}. Nevertheless, zero-area pulses, characterized by alternating signs and a vanishing time integral, are commonly employed and offer a promising route to high-fidelity control~\cite{PE,LM}. The analytical control conditions for such pulses have been derived using the Feshbach PQ partitioning technique~\cite{WAEST,LEO}. These pulses have been applied to various quantum information processing tasks~\cite{DI}, including state transfer~\cite{PC}, adiabatic quantum computing~\cite{EA,AA}, effective expedited holonomic quantum computation~\cite{PE}, and quantum thermodynamic processes~\cite{KT,WN,NZ,WG}. This framework, however, relies on the rapid oscillatory averaging, which requires the driving intensity to be much larger than the characteristic strength of the system. For instance, the results in Ref.~\cite{WAEST} require the pulse strength to be roughly 50 times or more the system's characteristic strength. Experimentally, the driving intensity is constrained by physical limits: in superconducting qubit systems, exceeding a threshold driving frequency can trigger an uncontrollable ionization effect, leading to a complete loss of quantum information~\cite{xia2025}. Theoretically, strong driving can introduce non-Markovian corrections by modifying the system-environment coupling, distorting the pulse waveform~\cite{dann2025}.

To design control pulses beyond the high-frequency limit, one needs a framework that can incorporate the effect of finite driving intensity. The Magnus expansion provides such a framework. It expresses the time-evolution operator as \(U(t)=\exp[-iH_{\rm eff}(t)]\), where \(H_{\rm eff}(t)\) is built from nested commutators \cite{Magnus1954,BLANES2009151}. This expansion remains unitary when truncated at any order \cite{Casas2001FloquetTE}, making it particularly suitable for quantum control. In dynamical decoupling, the first-order Magnus term has been used to derive DD conditions \cite{osti_321990}, while higher-order terms have been employed to estimate residual errors \cite{PhysRevA.84.012305}. As a general framework for time-dependent Hamiltonians, the Magnus expansion has been widely applied to closed and open quantum systems \cite{dai2016} and to the design of high-fidelity control pulses \cite{di2021}. When the driving frequency becomes comparable to the characteristic frequency of the system, however, the higher-order Magnus terms neglected in the high-frequency approximation need to be taken into account, as they can break the original control conditions and degrade the leakage suppression efficiency.

In this work, we have extended LEO-based quantum control beyond the high-frequency driving limit by reformulating it within the Magnus framework. This provides a systematic way to incorporate higher-order corrections when the driving frequency becomes comparable to the system energy scale. We will use near-perfect state transfer through a one-dimensional spin chain as an example to show how higher-order corrections improve the transmission fidelity at low driving intensities.

\section{Pulse conditions derived by the Feshbach PQ partitioning framework}

In this section, we will first introduce the construction of the LEO Hamiltonian, which can effectively prevent the leakage from the target subspace to the leakage subspace. Then we will derive the analytical pulse control conditions via Feshbach PQ technique.
Suppose there are two complete orthonormal basis $|\psi_n(0)\rangle$ and $|\psi_n(t)\rangle$, with $\langle \psi_m(0) |\psi_n(0)\rangle=\delta_{mn}(0)$, $\langle \psi_m(t) |\psi_n(t)\rangle=\delta_{mn}(t)$. There is a one to one correspondence between the states $|\psi_n(0)\rangle$ and $|\psi_n(t)\rangle$. An LEO Hamiltonian $H_{LEO}(t)$ has been proposed to eliminate the leakage of the states from and the encoded subspace $|\psi_0(t)\rangle$ to other subspaces~\cite{WM}. Now the total Hamiltonian is written as
\begin{eqnarray}
H(t)=H_{0}(t) + H_{LEO}(t),
\label{eq1}
\end{eqnarray}
where $H_0(t)$ is the original Hamiltonian.
\begin{eqnarray}
H_{LEO}(t)=c(t)A(t),
\label{eq10}
\end{eqnarray}
with $A(t)=|\psi_0(t)\rangle\langle\psi_0(t)|$. $c(t)$ is the control function that describes a sequence of control pulses.

Now we derive the pulse control conditions by using Feshbach PQ partitioning technique~\cite{Wunotes}. Considering the time dependent Hamiltonian $H_0(t)$, the system dynamics is given by the corresponding Schrödinger equation,
\begin{eqnarray}
i|\dot{\Psi}(t)\rangle = H(t)|\Psi(t)\rangle.
\end{eqnarray}
The state vector $|\Psi(t)\rangle$ can be expanded in terms of the basis $|\psi_n(t)\rangle$ and substituting it into Eq.~\eqref{eq1} yields
\begin{eqnarray}
\label{Eq.(3)}
i\dot{a}_{n} = \sum_{m} \left[\langle\psi_{n}|H(t)|\psi_{m}\rangle - i\langle\psi_{n}|\dot{\psi}_{m}\rangle\right] a_{m}.
\end{eqnarray}
The right side of Eq.~\eqref{Eq.(3)} is the effective Hamiltonian
\begin{eqnarray}
[\tilde{H}_{\rm eff}(t)]_{nm} = \langle\psi_{n}(t)|H(t)|\psi_{m}(t)\rangle - i\langle\psi_{n}(t)|\dot{\psi}_{m}(t)\rangle.
\label{eq04}
\end{eqnarray}
which is equivalent to the gauge transformation from the time independent basis $|\psi_n(0)\rangle$ to the time dependent basis $|\psi_n(t)\rangle$,
\begin{eqnarray}
\tilde{H}_{\rm eff}(t)=V_1^{\dagger}(t)H(t)V_1(t)-iV_1^\dagger(t)\partial_t V_1(t),
\label{eq05}
\end{eqnarray}
with $V_1(t)=\sum_n|\psi_n(t)\rangle\langle\psi_n(0)|$.

To keep the state in the subspace of $|\psi_0(t)\rangle$, one can employ the Feshbach PQ partitioning technique~\cite{LEO} of decomposing the n-dimensional state vector $|\Psi(t)\rangle$ into one dimensional state vector $P$ and the $(n-1)$-dimensional state vector $Q$ to obtain the precise one-component dynamical equation for the coefficient before $|\psi_0(t)\rangle$. Therefore, the Hamiltonian is divided into three parts, including $H\textsubscript{P}(t)$ and $H\textsubscript{Q}(t)$ acting on the subspace defined by $P$ and $Q$, and the remaining off-diagonal terms contributions
$H_0(t) = H_{P}(t) + H_{Q}(t) + H_{L}(t)$
\begin{eqnarray}
|\Psi(t)\rangle = \begin{bmatrix} P(t) \\ Q(t) \end{bmatrix}, \quad H_0(t) = \begin{bmatrix} h(t) & R(t) \\ W(t) & D(t) \end{bmatrix},
\label{PQmatrix}
\end{eqnarray}
Note that adding LEO in $H_0(t)$ is equivalent to $h(t) \rightarrow h^{'}(t)=h(t)+c(t)$ in $H_0(t)$. $P$ is the one dimensional contribution and the component $p(t)$ related to it can be expressed as
\begin{eqnarray}
p(t) = \exp\left[-i\int_{0}^{t} h^{'}(s') ds'\right] P(t).
\end{eqnarray}
In the selected one dimensional subspace, the projected Schrödinger equation becomes
\begin{eqnarray}
\dot{p}(t) = \int_{0}^{t} \tilde{g}(t,s) p(s) ds,
\end{eqnarray}
where the propagator $\tilde{g}(t,s)$ is given by
\begin{eqnarray}
\begin{aligned}
\tilde{g}(t,s) = -R(t)\mathcal{T}\left\{ \exp\left[-i\int_{s}^{t} D(s') ds'\right] \right\}\\
W(s)\exp\left[-i\int_{s}^{t} h^{'}(s') ds'\right].
\end{aligned}
\end{eqnarray}
Here $\mathcal{T}$ is the time ordering operator and
\begin{eqnarray}
\label{Eq.(9)}
\dot{p}(t) = -\int_{0}^{t} g(t,s) e^{-i\int_{s}^{t} h^{'}(s')ds'}p(s) ds.
\end{eqnarray}

\begin{eqnarray}
\tilde{g}(t,s)=-g(t,s)\exp[i\int_s^th(s')ds'],
\end{eqnarray}
In general, the above integral-differential equation has to be numerically solved. Now in the strong driving regime ($c(t) \gg h(t)$), i.e., $h^{'}(s)$ in Eq.~\eqref{Eq.(9)} is dominated by $c(s)$, we can first take $p(s) \rightarrow p(t)$, i.e., the wave function varies slowly compared to the driving. Secondly, $g(t,s)$, which is often reduced to $g(t-s)$, is a bath correlation function and often has common behaviors that hardly depend on the details of baths in the leakage subspace~\cite{PhysRevA.86.032303}. Normally, the real part of the function $g(t-s)$ starts from one and decays (rapidly) to zero with time $t-s$, unless baths have limited numbers of frequencies. In the strong driving case, $g(t,s) \rightarrow g(t,t)$ is also reasonable. Then $g(t,s)p(s)$ changes slowly compared to $\exp\left[-i\int_{s}^{t}c(s')ds'\right]$, Eq.~\eqref{Eq.(9)} simplifies to
\begin{eqnarray}
\label{Eq.(10)}
\int_{0}^{\tau} ds \exp\left[-i\int_{0}^{s} c(s') ds'\right] = 0.
\end{eqnarray}
 in a single control time interval $\tau$. Take the periodic rectangular pulses and sinusoidal pulses (also called zero area pulse in Ref.~\cite{LEO}) as examples to discuss the control conditions. The rectangular pulses can be set as 
\begin{eqnarray}
\label{eqmatrix}
c(t)=\begin{cases}
	I, n\tau<t<(n+1)\tau,\text{ for } n \text{ even}, \\
	-I,  \text{otherwise},
\end{cases}
\end{eqnarray}
here $I$ is the  strength of the control pulses, while $\tau$ is the half period of the pulses. By inserting Eq.~\eqref{eqmatrix} into Eq.~\eqref{Eq.(10)} one obtains~\cite{WAEST}
\begin{eqnarray}
\label{Eq.(12)}
I\tau = 2m\pi \quad \text{for} \quad m = 1,2,3,\ldots.
\end{eqnarray}

For the sinusoidal pulse, 
\begin{eqnarray}
\label{Eqsin}
c(t)=I\sin(\omega t),
\end{eqnarray}
 where $I$ is the pulse amplitude. One can have~\cite{WAEST}
\begin{eqnarray}
\label{Eq.(14)}
J_{0}\left(\frac{I\tau}{\pi}\right)=0,
\end{eqnarray}
where $J_{0}(x)$ is the zero order Bessel function of the first kind, and $\omega\tau=\pi$.

\section{Pulse control conditions derived by Magnus framework}

To apply the Magnus expansion, we first transform to a rotating frame via a gauge transformation that absorbs the rapid phase of the control pulse. 
Apply the gauge transformation \(V_2(t)\) to the effective Hamiltonian \(\tilde{H}_{\rm eff}\) in Eq.~(\ref{eq04}),
\begin{eqnarray}
\label{Po}
V_2(t)=\exp{\left[-i\phi(t){G}\right]}.
\end{eqnarray}
Then the effective Hamiltonian $H_{\rm eff}$ can be obtained
\begin{eqnarray}
\label{Po0}
H_{\rm eff}(t)=V_2^{\dagger}(t)\tilde{H}_{\rm eff}(t)V_2(t)-iV_2^\dagger(t)\partial_t V_2(t),
\end{eqnarray}
Here $\phi(t)=\int_{0}^{t}c(t')dt'$, ${G}$ is the generating operator. We move to a coordinate system that rotates with the integrated pulse phase $\phi(t)$. In this frame, the rapid oscillations of the control pulse are absorbed into the time evolution of the basis states $|\psi_n(t)\rangle$, leaving an effective Hamiltonian $H_{\rm eff}$ that contains only slowly varying couplings. As a result, the Feshbach PQ conditions, which were originally valid only under the high frequency approximation (corresponding to the zeroth order Magnus term), can be naturally extended to the low frequency driving regime by incorporating higher order terms.

After the gauge transformation \(V_2(t)\), the effective Hamiltonian in the P-Q block form in Eq.~(\ref{PQmatrix}) becomes
\begin{eqnarray}
H_{\rm eff}(t)=
\begin{pmatrix}
h(t) & e^{i\phi(t)}R(t)\\
e^{-i\phi(t)}R^{\dagger}(t) & D(t)
\end{pmatrix}.
\end{eqnarray}
To eliminate the diagonal dynamics, we define \(U_D(t)\) and \(\eta(t)\) by
\(i\dot{U}_D(t)=D(t)U_D(t)\) and
\(\eta(t)=\exp[-i\int_0^t h(s)ds]\), respectively, and perform the interaction-picture transformation
\(W(t)=\mathrm{diag}[\eta(t),U_D(t)]\).
Then the transformed Hamiltonian becomes
\begin{eqnarray}
H_{\rm eff}(t)=
\begin{pmatrix}
0 & v^{\dagger}(t)\\
v(t) & 0
\end{pmatrix},
\end{eqnarray}
with
\begin{eqnarray}
v(t)=e^{-i\phi(t)}U_D^{\dagger}(t)R(t)\eta(t).
\end{eqnarray}
This transformation removes the diagonal blocks \(h(t)\) and \(D(t)\), so that the leading Magnus terms are governed solely by the off-diagonal coupling \(v(t)\).

The general form of the Magnus expansion is~\cite{ebrahimi2023}
\begin{eqnarray}
\begin{aligned}
&H_{\rm eff}^{(0)}=\frac{1}{t}\int_{0}^{t} H_{\rm eff} dt,\\
&H_{\rm eff}^{(1)}=\frac{1}{2it}\int_{0}^{t}dt_1\int_{0}^{t_1}dt_2[{H}_{\rm eff}(t_1), H_{\rm eff}(t_2)],\\
&H_{\rm eff}^{(2)} =-\frac{1}{6t} \int_0^t dt_1 \int_0^{t_1} dt_2 \int_0^{t_2} dt_3 \\
&\quad \times \Big( [H_{\rm eff}(t_1), [H_{\rm eff}(t_2), H_{\rm eff}(t_3)]] \\
&\quad \quad + [H_{\rm eff}(t_3), [H_{\rm eff}(t_2), H_{\rm eff}(t_1)]] \Big)...
\end{aligned}
\end{eqnarray}

 The control condition requires that the off-diagonal matrix elements vanish, i.e.,
\begin{eqnarray}
\langle \psi_n | \sum_{k=0}^{\infty} H_{\rm eff}^{(k)} | \psi_m \rangle = 0,
\quad n \neq m.
\label{eq:001}
\end{eqnarray}

For the zeroth-order term, 
\begin{eqnarray}
\langle\psi_n|H_{\rm eff}^{(0)}|\psi_m\rangle=\frac{g_{mn}}{t}\int_0^t e^{i\phi(t)}dt=0.
\end{eqnarray}
Here, we use $g_{mn}$ to represent the interaction of internal couplings within the system. That is, in the off-diagonal matrix elements, the transformed interaction terms. For the rectangular pulses \eqref{eqmatrix} and sinusoidal pulses \eqref{Eqsin}, the ideal pulse conditions recover Eq.~\eqref{Eq.(12)} and Eq.~\eqref{Eq.(14)}. This shows that the high-frequency driving approximation used in Eq.~\eqref{Eq.(10)} is exactly the zeroth-order Magnus condition.  While this zeroth-order condition is universal in the high-frequency limit, higher-order Magnus terms depend on the detailed level structure of the system and become non-negligible when the driving frequency is comparable to the system energy scale.

\section{near perfect state transfer through a spin chain}
As an application of the Magnus framework, we now study near-perfect state transfer in a one-dimensional $XY$ spin chain. This problem has been studied previously using the PQ partitioning technique~\cite{WAEST}, and here we revisit it to examine the higher-order Magnus corrections in the low-frequency driving regime. The task is to achieve near-perfect state transfer by adding an LEO Hamiltonian to the chain. The Hamiltonian of the chain is
\begin{eqnarray}
H_s=\sum_{i=1}^{N-1}J_{i,i+1}\left(\sigma_i^x \sigma_{i+1}^x + \sigma_i^y \sigma_{i+1}^y \right),
\end{eqnarray}
where $J_{i,i+1}$ is the nearest-neighbor coupling. For uniform couplings, $J_{i,i+1}=J$ is a constant.

For simplicity, our task is to transfer the state $|1\rangle$ from the first to the last spin of the chain, with the initial state $|100\cdots0\rangle$ and target state $|\textbf{N}\rangle = |000\cdots1\rangle$. During this process, the transmission fidelity $F(t) = |\langle \textbf{N}|\psi(t)\rangle|^2$ is monitored to evaluate the transfer quality. $J_{i,i+1}=\sqrt{i(N-i)}$ is defined as perfect state transfer (PST) couplings and the Hamiltonian is $H_{PST}$. In this case, PST can be realized~\cite{WAEST}. As in Ref.~\cite{WAEST}, $|\psi_0(t)\rangle$ is set as $\exp{(-iH_{PST}t)}|\textbf{1}\rangle$. Then the dynamics is limited in a single-excitation subspace,  in which the system can be mapped to a tight-binding model via the Jordan-Wigner transformation. Taking
\begin{eqnarray}
V_1(t)=e^{-iH_{PST}t},
\end{eqnarray}
and
\begin{eqnarray}
\begin{aligned}
V_2(t)=&\exp\left(-i\int_{0}^{t}c(t')dt'\,|1\rangle\langle1|\right)
\\&=\exp\left(-i\phi(t)|1\rangle\langle1|\right),
\end{aligned}
\end{eqnarray}
The interaction-picture transformation is
\begin{eqnarray}
W(t)=
\begin{pmatrix}
e^{-i\int_0^t h(s)\,ds} & 0\\
0 & e^{-i2JBt}
\end{pmatrix}.
\end{eqnarray}
So the off-diagonal element is
\begin{eqnarray}
v(t)=e^{i2JBt}\,e^{-i\phi(t)}\,R(t)\,e^{-i\int_0^t h(s)\,ds},
\end{eqnarray}
where $B$ satisfies 
\begin{eqnarray}
e^{-i2JBt}=\Sigma_ke^{-i2J\lambda_k t}|\lambda_k\rangle\langle\lambda_k|.
\end{eqnarray}
\(B\) represents the matrix of the spin-chain Hamiltonian in the Q subspace, and \(\lambda_k\) and \(|\lambda_k\rangle\) denote its eigenvalues and eigenstates. For the present case, the eigenvalues in the Q subspace are
\(\lambda_0=0,\ \lambda_\pm=\pm\sqrt{2}J\).

For the first-order Magnus term, we need to evaluate
\begin{eqnarray}
H_{\rm eff}^{(1)}
=\frac{1}{2it}\int_0^t dt_1\int_0^{t_1}dt_2\,
[H_{\rm eff}(t_1),H_{\rm eff}(t_2)],
\label{eq31}
\end{eqnarray}
where the effective Hamiltonian in the interaction picture takes the off-diagonal form
\begin{eqnarray}
H_{\rm eff}(t)=
\begin{pmatrix}
0 & v^{\dagger}(t)\\
v(t) & 0
\end{pmatrix}.
\end{eqnarray}
The commutator in the right side of Eq.~(\ref{eq31}) is
\begin{eqnarray}
\begin{pmatrix}
v^{\dagger}(t_1)v(t_2)-v^{\dagger}(t_2)v(t_1) & 0\\
0 & v(t_1)v^{\dagger}(t_2)-v(t_2)v^{\dagger}(t_1)
\end{pmatrix},
\end{eqnarray}
which is also diagonal. Therefore, after integration, \(H_{\rm eff}^{(1)}\) remains diagonal, and its off-diagonal matrix elements vanish identically,
\begin{eqnarray}
\langle\psi_n|H_{\rm eff}^{(1)}|\psi_m\rangle=0,\qquad n\neq m.
\end{eqnarray}
Thus the first-order Magnus term only shifts the energies within the \(P\) and \(Q\) subspaces, it does not induce transitions or leakage between them. This structure persists at higher orders. Since \(H_{\rm eff}(t)\) is off-diagonal, its nested commutators
$[H_{\rm eff}(t_1),[H_{\rm eff}(t_2),\dots,[H_{\rm eff}(t_{n-1}),H_{\rm eff}(t_n)]\dots]]$ are off-diagonal for odd \(n\) and diagonal for even \(n\). Consequently, the odd-order Magnus terms contribute only to the diagonal blocks and cannot generate leakage. Leakage arises solely from the even-order terms. It should be noted, however, that the odd-order terms still modify the effective Hamiltonian in the \(Q\) subspace, which in turn affects the second-order off-diagonal elements and hence the leakage.

Since the higher-order leakage term is complex, it cannot be set to zero exactly. Instead, we minimize its squared modulus,
\begin{eqnarray}
\label{32}
\left|\langle\psi_n|H_{\rm eff}^{(0)}+H_{\rm eff}^{(2)}+\cdots|\psi_m\rangle\right|^2,
\end{eqnarray}
so that the leakage is as small as possible.

Based on the Magnus expansion, we obtain an empirical formula in the low-frequency regime,
\begin{eqnarray}
\theta \approx j_{0,k}-\frac{2\tau}{\pi}J_1(j_{0,k}),\qquad I\le 12,
\label{eq35}
\end{eqnarray}
where \(\theta=I\tau/\pi\) and \(j_{0,k}\) is the \(k\)-th zero of \(J_0(x)\).
Therefore, the half-period $\tau$ corresponding to the determined pulse driving intensity is
\begin{eqnarray}
\label{37}
\tau\approx \frac{j_{0,k}\pi}{I+2J_1(j_{0,k})},I\leq12.
\end{eqnarray}

Finally, for low-frequency driving, there is a broad sufficient condition for the convergence of the Magnus expansion~\cite{Casas_2007},
\begin{eqnarray}
\int_0^t ||H_{\rm eff}(t)||dt<\pi.
\end{eqnarray}
This condition implies that the driving intensity cannot be too weak; otherwise the expansion may fail to converge. In our case, this gives a lower bound around \(I\ge 3.8\). However, this bound is only sufficient, not necessary. For instance, in the range \(3\le I\le 4\), the integral can exceed \(\pi\) while the expansion still converges.

\begin{figure}[htbp]
\centering
\begin{minipage}{0.45\textwidth}
\centering
\textbf{(a)}\\[2pt]
\includegraphics[width=\textwidth]{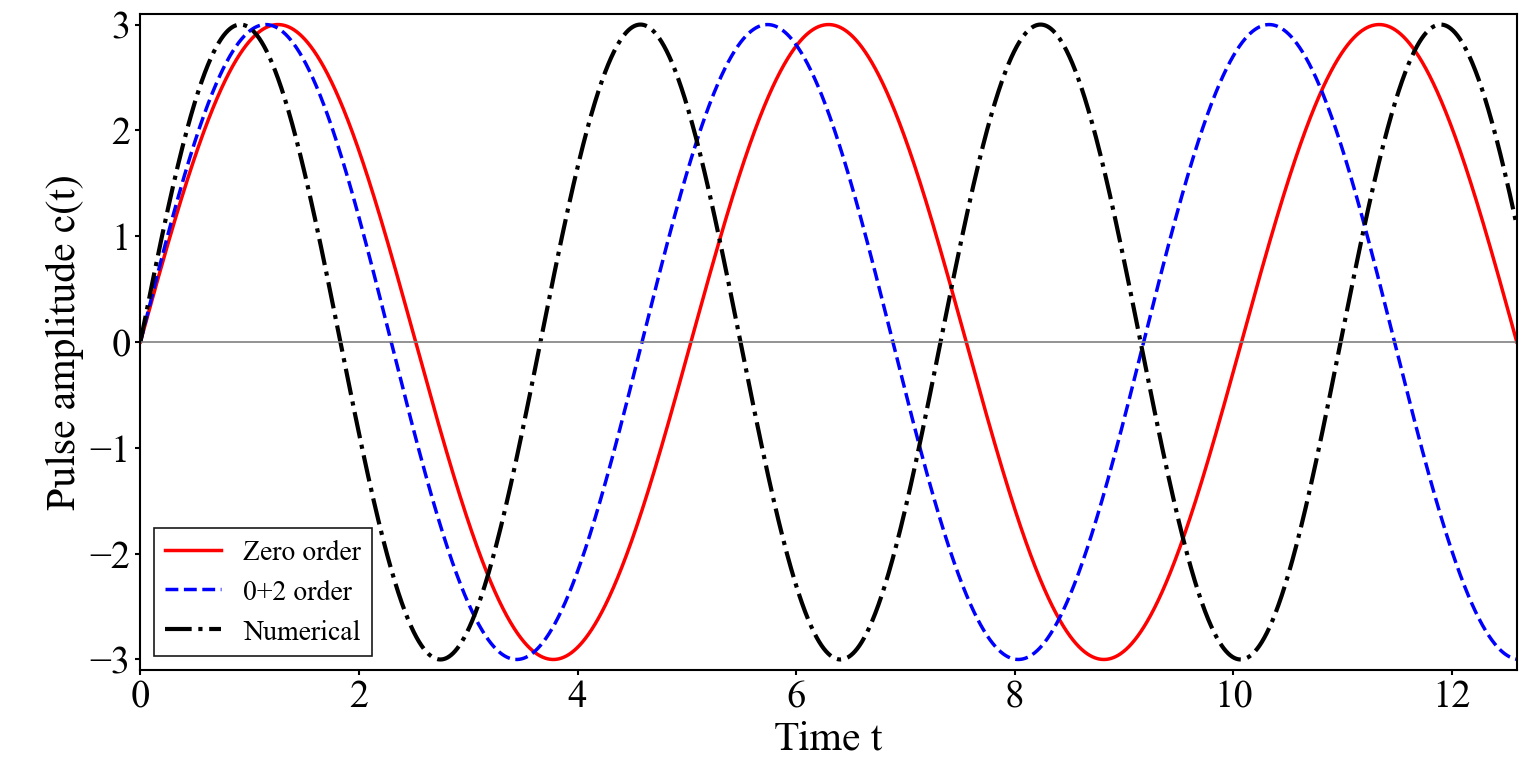}
\label{fig:1a}
\end{minipage}
\hfill
\begin{minipage}{0.45\textwidth}
\centering
\textbf{(b)}\\[2pt]
\includegraphics[width=\textwidth]{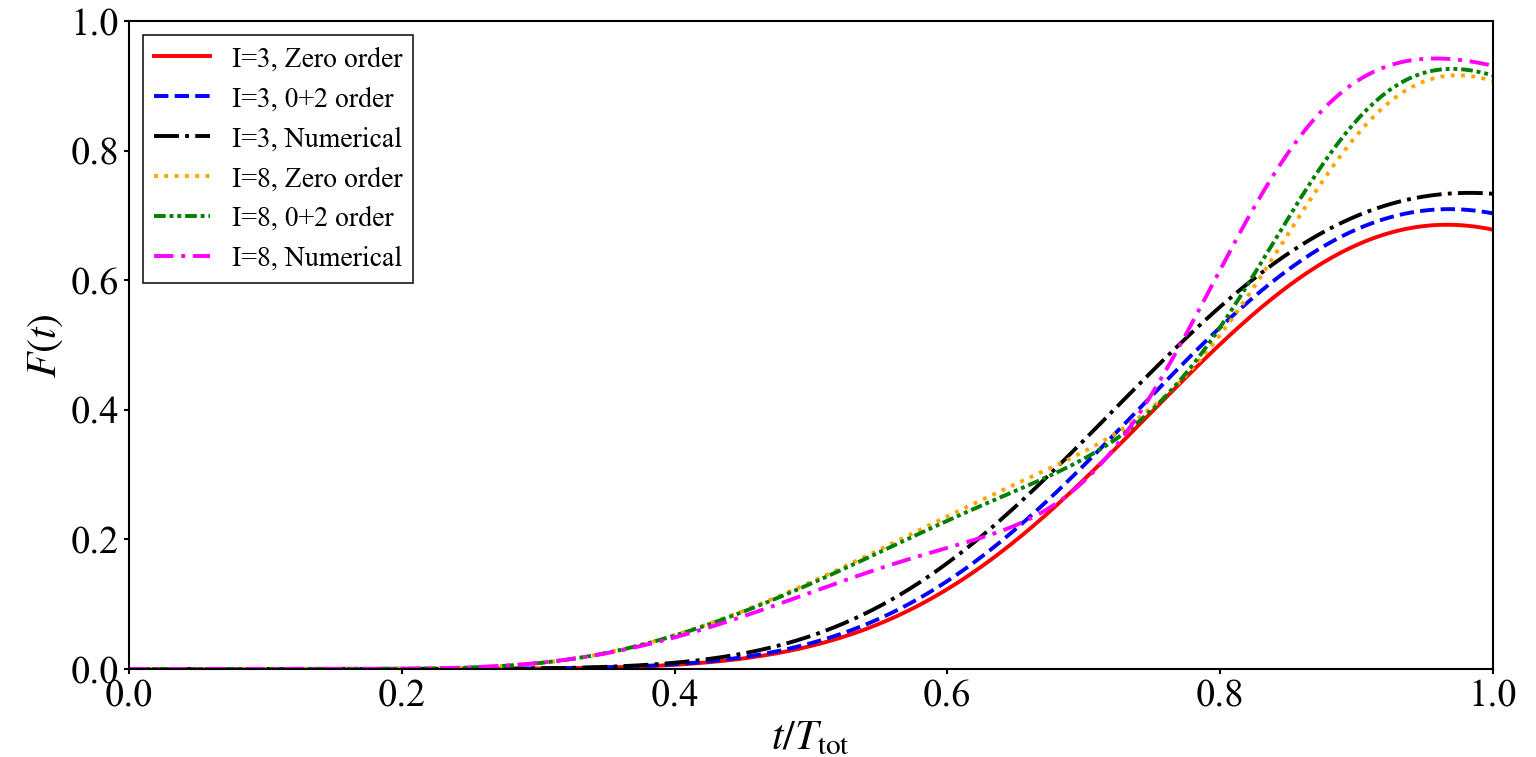}
\label{fig:1b}
\end{minipage}
\caption{(a) The control functions \(c(t)\) from the zeroth-order, 0+2-order, and numerically optimized conditions at \(I=3\); (b) The transmission fidelity \(F\) versus the rescaled evolution time \(t/T_{\rm tot}\) for \(N=4\) and \(T_{\rm tot}=\pi/2\).}
\label{fig:1}
\end{figure}

\begin{figure}[htbp]
\centering
\includegraphics[width=0.45\textwidth]{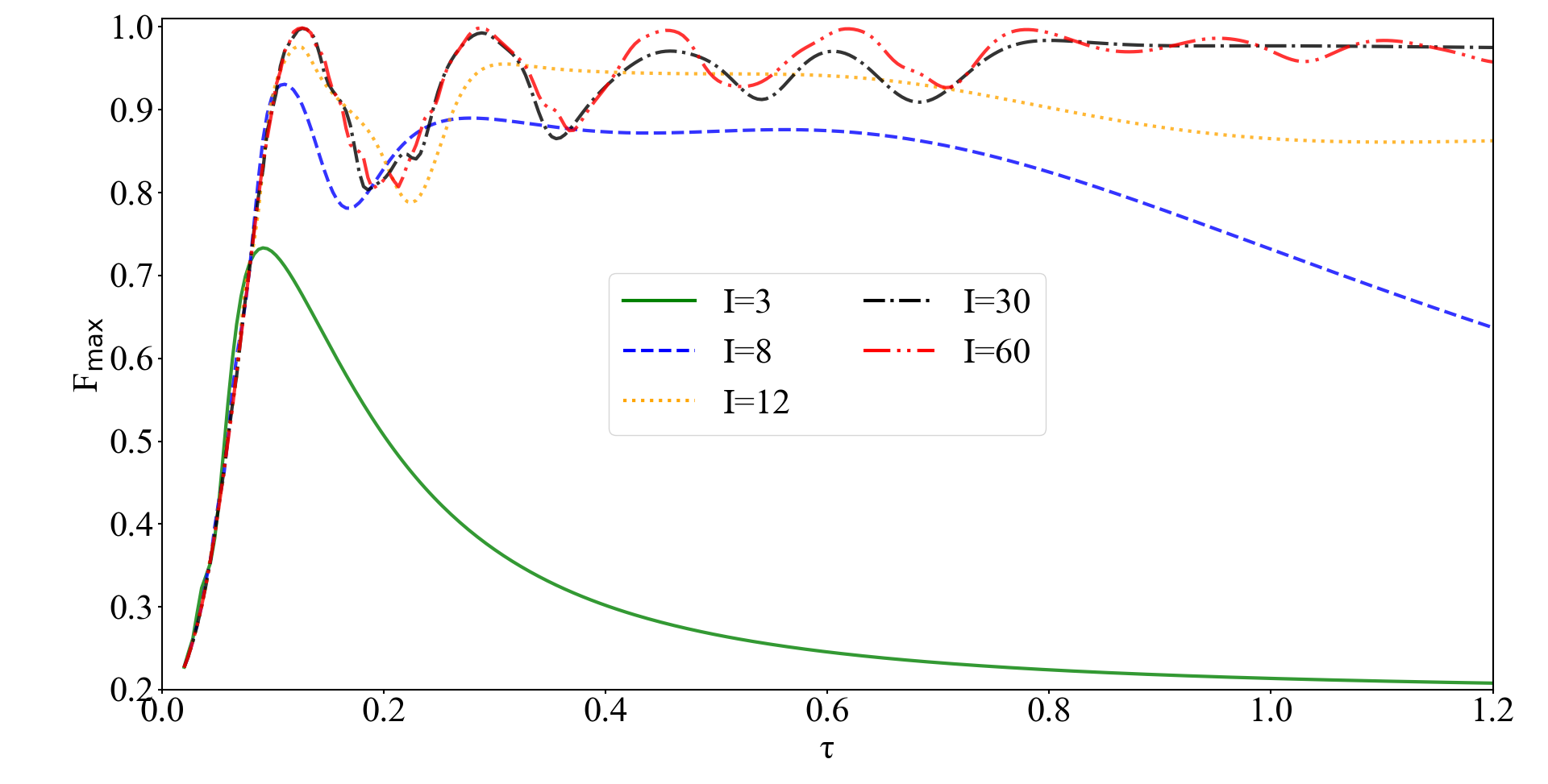}
\caption{The maximum fidelity $F_{max}$ versus the pulse half period $\tau$ for different pulse intensity $I$. $N=4$.}
\label{fig:2}
\end{figure}

Fig.~\ref{fig:1}(a) compares the control functions obtained from the zeroth-order, 0+2-order, and numerically optimized conditions at \(I=3\). For the zeroth-order condition \(J_0(I\tau/\pi)=0\), we take the first root, \(I\tau/\pi=2.4048\). The 0+2-order condition shifts this root to \(I\tau/\pi\approx2.1912\). This leftward shift at low frequency is clearly visible.

Fig.~\ref{fig:1}(b) shows the transmission fidelity \(F\) versus the rescaled evolution time \(t/T_{\rm tot}\) for different \(I\), with \(N=4\) and \(T_{\rm tot}=\pi/2\). At \(I=3\), the 0+2-order approximation improves the fidelity appreciably over the zeroth-order result, indicating that the zeroth-order condition becomes insufficient as the driving frequency decreases and higher-order Magnus terms must be included. As \(I\) increases from 3 to 8, the 0+2-order fidelity approaches the zeroth-order result, and in this case the zeroth-order description is sufficient.

In Fig.~\ref{fig:2}, we plot the maximum fidelity \(F_{\max}\) as a function of the pulse half-period \(\tau\) for different driving intensities \(I\), obtained by direct numerical search without imposing any control condition. Here \(N=4\), and \(\tau\) is rescaled as \(I\tau/60\), with \(60\) being the largest driving intensity considered, so that the peaks for different \(I\) align on a common scale. Otherwise, the peaks for small \(I\) would fall outside the plotted range.

For high-frequency driving, multiple peaks are observed. Their positions are consistent with the roots of the zeroth-order condition \(J_0(I\tau/\pi)=0\), confirming that the numerical optima follow the theoretical pulse conditions. In contrast, for low-frequency driving, only a few peaks are resolved. This is because, for a fixed total evolution time, a larger \(\tau\) corresponds to a longer pulse period, so fewer complete pulse cycles fit into the evolution.

\begin{figure}[htbp]
\centering
\includegraphics[width=0.45\textwidth]{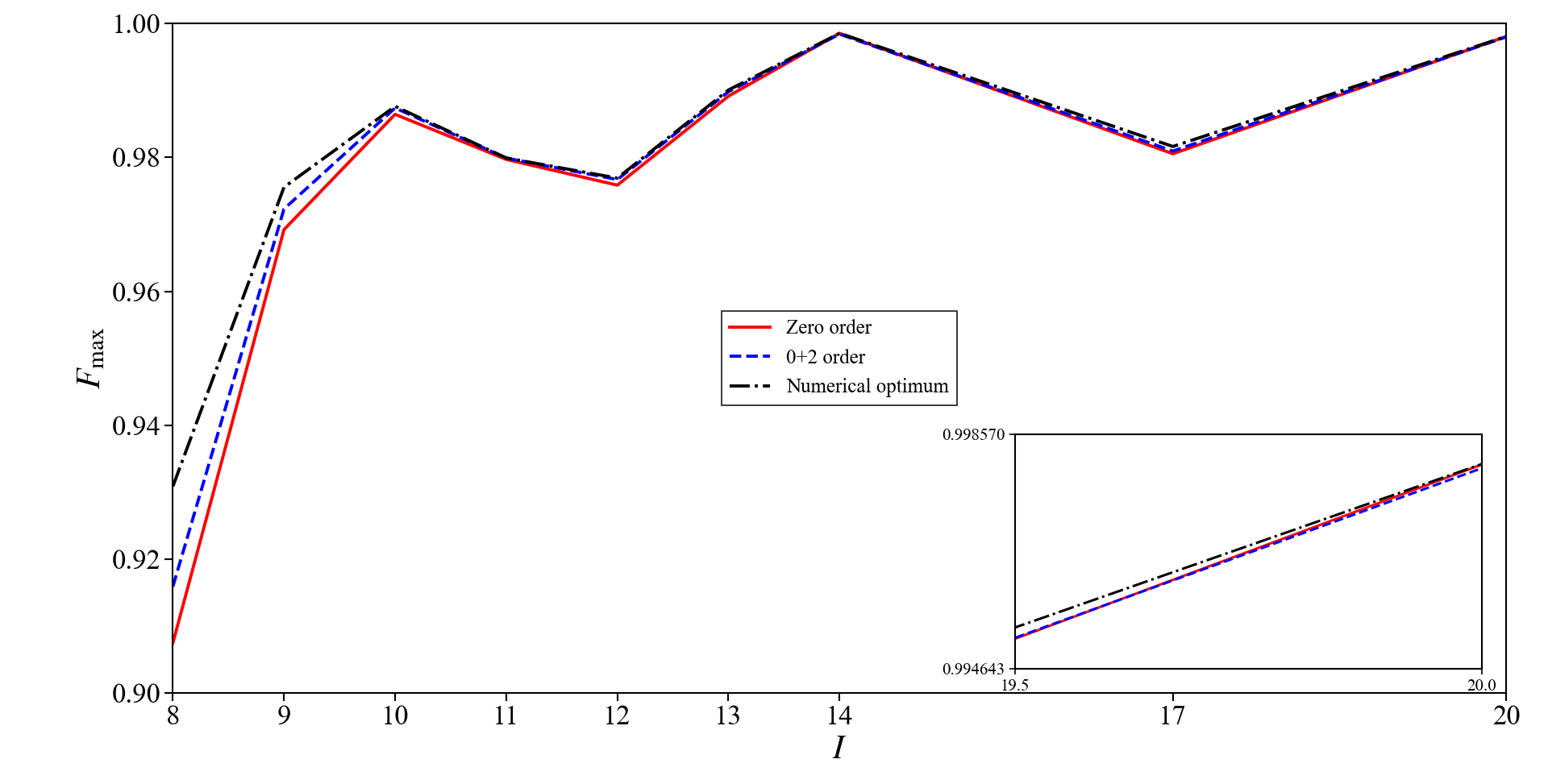}
\caption{The first-peak fidelity versus pulse intensity for the zeroth-order, 0+2-order, and numerical solutions.}
\label{fig:3}
\end{figure}

We now examine the first peak for each \(I\) to assess the effect of higher-order corrections on the transmission fidelity. Fig.~\ref{fig:3} compares the maximum fidelities obtained from the numerical solution, the zeroth-order condition, and the 0+2-order condition for various \(I\). In the low-frequency regime, the 0+2-order result approaches the numerical solution more closely than the zeroth-order one, whereas in the high-frequency regime the zeroth-order approximation is sufficiently accurate. For \(I\le20\), the 0+2-order fidelity is consistently slightly higher than the zeroth-order one. For example, at \(I=8\), the 0+2-order fidelity reaches \(F_{\max}=0.916\), compared with \(F_{\max}=0.907\) for the zeroth-order condition. This confirms that higher-order corrections provide an improvement at low-frequency driving, while their effect diminishes as \(I\) increases.

\begin{figure}[htbp]
\centering
\includegraphics[width=0.45\textwidth]{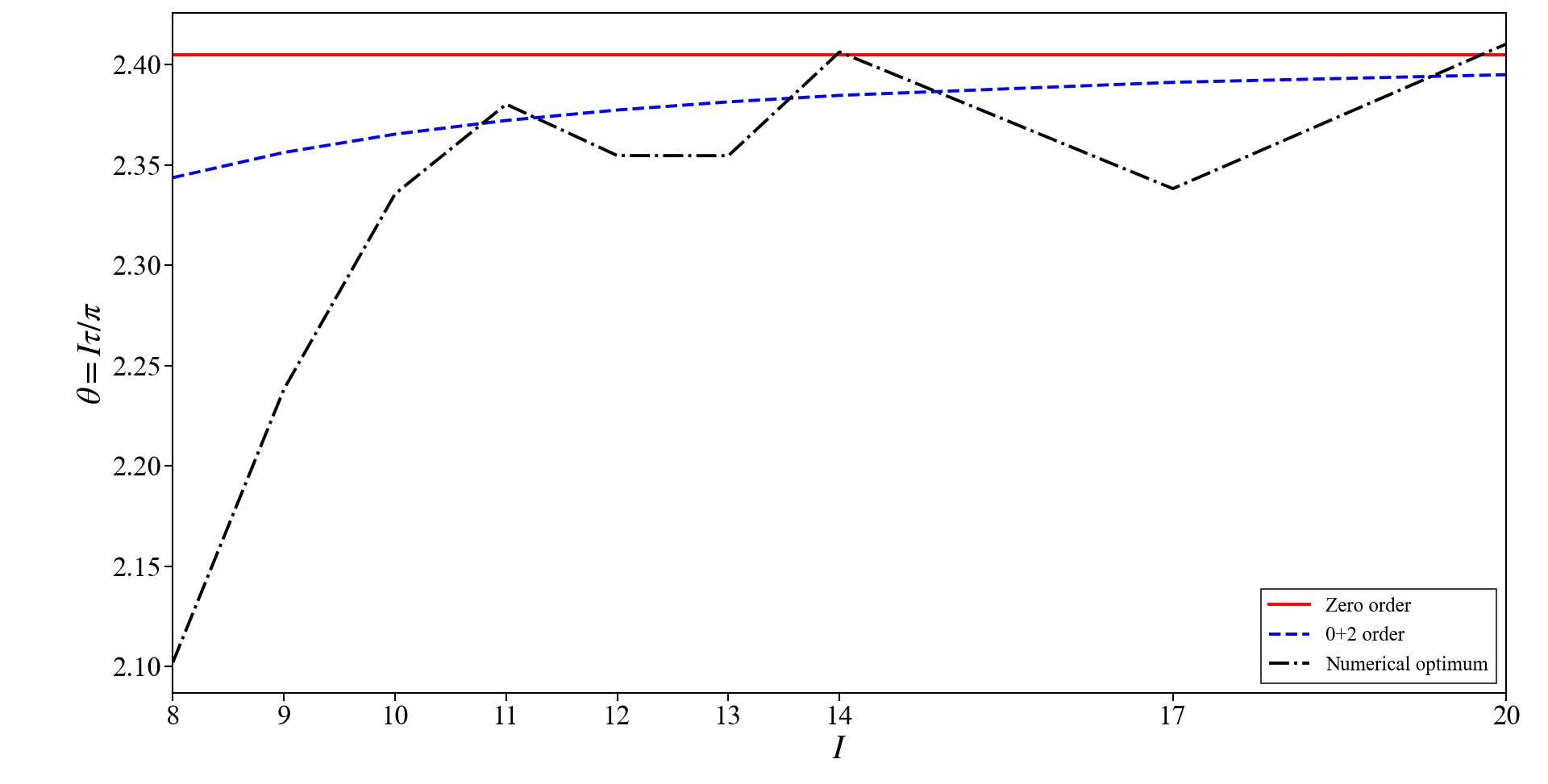}
\caption{\(I\tau/\pi\) versus pulse intensity \(I\) for the zeroth-order, 0+2-order, and numerically optimized conditions in the low-frequency regime.}
\label{fig:4}
\end{figure}

Fig.~\ref{fig:4} shows \(I\tau/\pi\) in the low-frequency regime for the zeroth-order, 0+2-order, and numerically optimized conditions. The zeroth-order condition gives a constant \(I\tau/\pi=j_{0,k}\), whereas the 0+2-order and numerical results shift to smaller values as \(I\) decreases. For example, at \(I=9\), the zeroth-order condition gives \(\tau=0.8394\) (\(I\tau/\pi=2.4048\)), the 0+2-order condition gives \(\tau=0.8225\) (\(I\tau/\pi=2.3562\)), and the numerical optimum is \(\tau=0.7813\) (\(I\tau/\pi=2.2382\)). This confirms that the higher-order correction moves the pulse condition toward the numerical optimum, corresponding to a leftward shift of the Bessel zero at low frequencies. The empirical Eq.(\ref{eq35}) reproduces this leftward shift in the low-frequency regime. After determining the low-frequency pulse intensity $I$, the corresponding half-period can also be obtained from Eq.(\ref{37}).

Overall, the results demonstrate that higher-order Magnus corrections improve the control performance in the low-frequency regime by shifting the zero of the Bessel function \(J_0\) leftward ~\cite{PhysRevLett.84.2112,PhysRevB.67.165301}, toward the numerically optimal value. At high frequencies, the improvement becomes negligible, since the zeroth-order condition is already sufficiently accurate. These results support the general framework developed in Sec.~III and clarify the regime in which higher-order corrections are beneficial for low-frequency LEO control.

\section*{Conclusions}

In this work, we reformulated LEO within the Magnus framework and extended it beyond the high-frequency driving limit. We found that the traditional high-frequency control conditions correspond to the zeroth-order Magnus approximation, while higher-order corrections become essential at low driving frequencies. This is accompanied by a leftward shift of the \(J_0\)-type zeros, consistent with previous observations~\cite{PhysRevLett.84.2112,PhysRevB.67.165301}. Using LEO-controlled near-perfect state transfer in a one-dimensional spin chain as an example, we demonstrated that the second-order Magnus correction improves the fidelity at low driving intensities. This result has potential implications for quantum control design in solid-state platforms, such as superconducting qubits, where the driving intensity is often constrained by hardware limits~\cite{Repicky2026Unified}. This framework can also be extended to open quantum systems in the future, in which the coupling between the system and its environment may induce additional effects.

\section*{acknowledgements}
We thank H.-R. Li and K. Zhou for helpful discussions. This paper is supported by the Natural Science Foundation of Shandong Province (Grant No. ZR2024MA046, ZR2026LLZ009) and the Fundamental Research Funds for the Central Universities (Grant No. 202661013) through the Graduate Independent Research Project of Ocean University of China.

\bibliography{ref}
\end{document}